# Topologically protected photonic transport in bi-anisotropic meta-waveguides

Tzuhsuan Ma[1], Alexander B. Khanikaev[3], S. Hossein Mousavi[2], Gennady Shvets[1,*]

[1] Department of Physics, The University of Texas at Austin, Austin, Texas 78712, USA
[2] Department of Electric Engineering, The University of Texas at Austin, Austin, Texas 78712, USA
[3] Department of Physics, Queens College of The City University of New York, Flushing, New York 11367, USA
and The Graduate Center of The City University of New York, New York, New York 10016, USA

*gena@physics.utexas.edu

We propose a design of two-dimensional bi-anisotropic meta-waveguide exhibiting topologically nontrivial photonic phase and robust photonic transport. The structure proposed represents a waveguide formed by a periodic array of metallic cylinders placed inside parallel metal plates and is designed to support two pairs of Dirac cones thus allowing emulation of the spin of the electrons in condensed matter systems. It is shown, that by making the meta-waveguide bi-anisotropic one can mimic the spin-orbital interaction in topological insulators and open a complete photonic topological band gap. First-principle numerical simulations demonstrate that the domain walls between two meta-waveguides with reversed bi-anisotropy support photonic edge states robust against certain classes of structural imperfections, including sharp bends and disorder.

Since the discovery of the Quantum Hall Effect (QHE) [1], topologically nontrivial phases of condensed matter systems have been of a significant interest for the scientific [2, 3, 4, 5, 6]. Realization of topological phases in other fundamentally different systems, including photonic ones, can also be of a great interest from the point of view of fundamental science, but can also potentially underpin new revolutionary concepts in applied physics and engineering. One of possible approaches to topologically nontrivial photonic states is to mimic their electronic counterparts by emulating spin of electrons and effects of magnetic field or spin-orbital interaction in photonic systems. This approach has been theoretically shown to be quite fruitful in a number of systems including magnetic photonic crystals [7, 8, 9, 10, 11], cavity arrays [12], coupled ring resonators [13], and bi-anisotropic metamaterials [14]. An alternative approach to engineer topological phases relying on temporal modulation of the photonic system has also been recently introduced. It was theoretically shown that this approach allows synthesizing an artificial gauge field for photons – synthetic magnetic field [15] giving rise to topologically protected transport of light. A similar concept of Floquet topological insulators, originally proposed for the condensed matter systems [16], has been recently brought to photonics [17]. Rechtsman and coauthors have theoretically predicted and experimentally demonstrated that by replacing temporal modulation with the spatial one it is possible to obtain a new class of topologically nontrivial of photonic states in coupled helical fibers [17].

In our recent work we have demonstrated that metamaterials [18, 19] represent another promising platform for emulation nontrivial photonic phase with preserved time-reversal symmetry [14] with bi-anisotropy [20, 21, 22, 23] playing the role analogous to spin-orbital coupling. In this letter, in order to bring this concept one step closer to experimental implementation, we propose a new simple design of two-dimensional bi-anisotropic structure with topological protection – bi-anisotropic meta-waveguide (BMW).

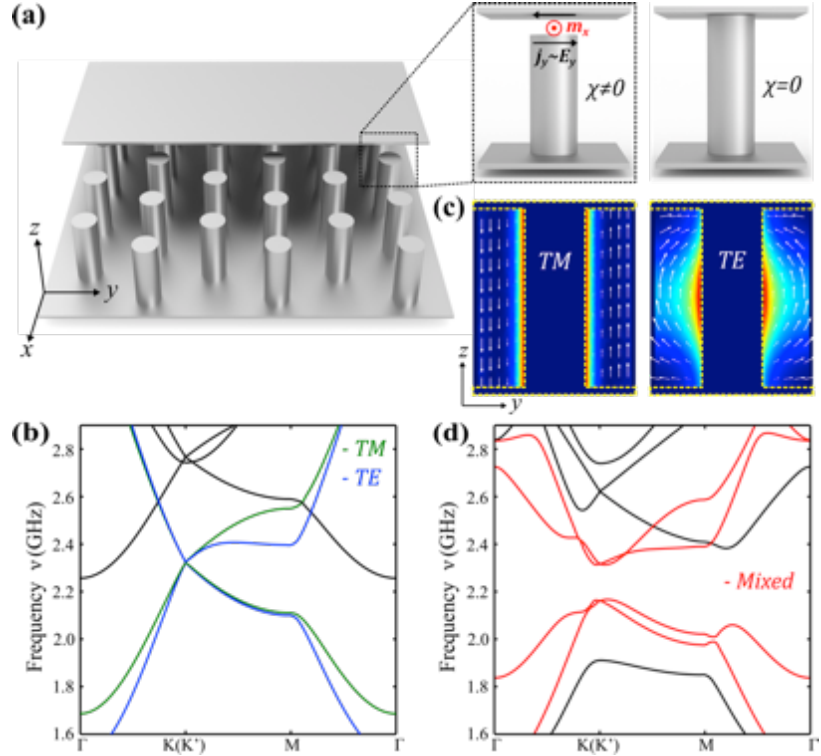

**Figure 1 (Color online)**. (a) Schematics of bi-anisotropic meta-waveguide with part of the top metal plate removed to see the structure below. The enlarged gap regions on the right illustrate bi-anisotropic response of the unit cell. (b) Photonics band structure (PBS) of spin-degenerate meta-waveguide with TE and TM modes forming doubly degenerate Dirac cones at $K(K')$ point (No bi-anisotropy present, $\chi = 0$). (c) Field profiles of the degenerate TE and TM mode at $K(K')$ point. The colors show the energy density. The arrows for TM profile are electric field, and the arrows for TE profile are magnetic field. The yellow dashed line outline where the metallic structures are, and there are no fields inside the metal. (d) PBS with the complete gap induced by the bi-anisotropy of the meta-waveguide ($\chi \neq 0$). The bands of interest are highlighted by color in (b) and (d). The structure parameters are as follows: the lattice period $a_0 = 10$ cm, distance between metal plates $h = 10$ cm, the cylinders' diameter $d = 3.45$ cm. For (b) and (d) the gap size is $g = 0$ and $g = 1.5$ cm, respectively.

Our approach to emulate the spin and spin-orbital interaction of electrons is based on two crucial steps outlined in Ref. [14]: (i) engineering of the spin-degeneracy, i.e. the photonic band structure with the two degenerate pairs of Dirac cones for the two polarizations of light, and (ii) introduction of bi-anisotropy mixing these polarizations in a specific way emulating spin-orbital coupling. In this letter we consider a meta-waveguide formed by the parallel-plate metal waveguide filled with a periodically arranged hexagonal array of metallic cylinders connecting the top and/or bottom metal plates, shown in Fig1 (a). Its modes can be classified as TE (with nonzero components $E_x, E_y, H_x, H_y, H_z$) and TM (with nonzero components $E_z, H_x, H_y$) polarized modes [24] while the hexagonal symmetry guarantees the presence of the Dirac cones in both polarization. In what follows we will focus on the lowest order Dirac points which appears between two dipolar. In the Bloch ansatz, the field profiles of TE and TM mode can be written as

$$\boldsymbol{E}_{TE}(\boldsymbol{r},t) = \boldsymbol{E}_{TE,\boldsymbol{k}_\perp}(\boldsymbol{r}_\perp)e^{i\boldsymbol{k}_\perp\cdot\boldsymbol{r}_\perp+i\frac{\pi}{h}z-i\omega t}, \quad (1.1)$$

$$\boldsymbol{E}_{TM}(\boldsymbol{r},t) = \boldsymbol{E}_{TM,\boldsymbol{k}_\perp}(\boldsymbol{r}_\perp)e^{i\boldsymbol{k}_\perp\cdot\boldsymbol{r}_\perp-i\omega t}, \quad (1.2)$$

where $\boldsymbol{k}_\perp = (k_x, k_y)$, $\boldsymbol{r}_\perp = (x,y)$, and $h$ is the distance between metal plates. In the following discussion, we focus on the fields in k-space.

To emulate the spin degree of freedom with the polarization of modes, we first implement the condition of "spin-degeneracy" [14], by ensuring the degeneracy between TE and TM modes achieved through varying parameters of the structure such as the separation between the metal plates and the filling fraction (radius) of the cylinders. As found by direct numerical calculations with the use of COMSOL Multiphysics, the condition of spin-degeneracy is met for the parameters given in the Fig. 1 caption. The resultant photonic band structure shown in Fig. 1(b) clearly shows degenerate TE and TM Dirac cones overlapping at the frequency of 2.32GHz. Note that the slope of the dispersion (the effective Dirac velocity $v_D = v_D^{TE} = v_D^{TM}$) for both TE and TM modes is identical in the vicinity of the Dirac point, which ensures efficient hybridization between the modes in the frequency range proximal to the Dirac point. This also allows description of the system in the proximity to $K$ and $K'$ points with the effective 8×8 Dirac Hamiltonian:

$$\mathcal{H}_0 = v_D \hat{s}_0 (\hat{\tau}_z \hat{\sigma}_x \delta k_x + \hat{\tau}_0 \hat{\sigma}_y \delta k_y), \quad (2)$$

Where $\delta k_x$ and $\delta k_y$ are deviations of the $x$ and $y$ components of the wave-vector $\boldsymbol{k}$ from $K(K')$-point, $\hat{\sigma}_i$, $\hat{\tau}_i$, and $\hat{s}_i$ are the Pauli matrices of the Dirac bands (corresponding to the two dipolar degrees of freedom), $K$ and $K'$ valleys, and polarization (spin) subspaces, respectively [14]. Note that since $v_D^{TE} = v_D^{TM}$ near $K$ and $K'$ points, any linear combination of TE and TM mode can be considered as an eigenmode of the system and the Hamiltonian has internal U(1) symmetry which is reflected in the Eq. (2) in only trivial presence of the spin degree of freedom through the identity matrix $\hat{s}_0$.

As the next step we introduce bi-anisotropy into the system by reducing the length of the cylinders such that a gap of the size $g$ opens *between the cylinders* and *one of the plates*. Presence of such gap breaks $\sigma_z$ mirror symmetry and brings in a bi-anisotropic response of the meta-waveguide [25]. The origin of the bi-anisotropy is schematically explained in Fig.1(a) (right insets) where one can see that the electric field in the gap between cylinder and the plate $E_{x(y)}$ induces anti-symmetric currents $j_{x(y)}$ producing the net orthogonal magnetic moment

$m_{y(x)}$, and vice versa. The result of the polarization hybridization on the PBS is shown in Fig. 1(c). A complete photonic band gap between 2.17 GHz and 2.31 GHz is formed.

The perturbation theory is also developed to quantitatively describe the relation between the deformation of the cylinder and the strength of the coupling between TE and TM modes (See supplementary part for more information). In the vicinity of the $K$ and $K'$ points, the mixed (perturbed) modes can be approximately expanded by just two modes, i.e. unperturbed TE and TM mode with analytic description Eq. (1). The degree of hybridization of the modes is proportional to the amount of deformation of the metal. According to the degenerate perturbation theory, the difference between the perturbed and unperturbed eigenvalues is given by Eq. (3) with $E'_{k_\perp}(r) = \sum_m a_m E_{m,k_\perp}(r)$, $H'_{k_\perp}(r) = \sum_m a_m H_{m,k_\perp}(r)$, and $m = TE$ or $TM$. Eq. (3) is written in frequency domain. The primed quantities are for the perturbed modes. $\omega$'s are the eigenfrequencies of the modes, and $a_m$'s are complex-valued coefficients representing the degree of hybridization.

$$\frac{\omega'(k_\perp) - \omega_m(k_\perp)}{\omega_m(k_\perp)} = -\frac{\int_{\Delta V}\left(\epsilon_0 E^*_{m,k_\perp} \cdot E'_{k_\perp} - \mu_0 H^*_{m,k_\perp} \cdot H'_{k_\perp}\right) dV}{\int_V \left(\epsilon_0 E^*_{m,k_\perp} \cdot E'_{k_\perp} + \mu_0 H^*_{m,k_\perp} \cdot H'_{k_\perp}\right) dV}, \qquad (3)$$

By expanding Eq. (3) with unperturbed modes, we recast it into an eigenvalue problem:

$$\begin{bmatrix} \omega_{TE}(1 + \Delta_{TE,TE}) & \omega_{TE}\Delta_{TE,TM} \\ \omega_{TM}\Delta_{TM,TE} & \omega_{TM}(1 + \Delta_{TM,TM}) \end{bmatrix} \begin{pmatrix} a_{TE} \\ a_{TM} \end{pmatrix} = \omega' \begin{pmatrix} a_{TE} \\ a_{TM} \end{pmatrix}, \qquad (4)$$

where $\Delta_{m,n} = -\int_{\Delta V}\left(\epsilon_0 E^*_{m,k_\perp} \cdot E_{n,k_\perp} - \mu_0 H^*_{m,k_\perp} \cdot H_{n,k_\perp}\right) dV$ with $\Delta V$ the deformed part of the cylinder, and $m, n = TE$ or $TM$. The dispersions of the interacting TE and TM mode degenerate at $K$ and $K'$ point, and have the group velocity with opposite sign, $\left(\frac{\partial \omega_{TE}}{\partial k_\perp}\right)_{K,K'} = -\left(\frac{\partial \omega_{TM}}{\partial k_\perp}\right)_{K,K'}$. Near $K$ and $K'$ point, the dispersion relations of TE and TM mode can be expressed as $\omega_{TE} = \omega_D + v_D \delta k$ and $\omega_{TM} = \omega_D - v_D \delta k$, or $\omega_{TE} = \omega_D - v_D \delta k$ and $\omega_{TM} = \omega_D + v_D \delta k$; where $\delta k = |\mathbf{k} - \mathbf{K}|$ or $|\mathbf{k} - \mathbf{K}'|$, $v_D$ is Dirac velocity, and $\omega_D$ is the crossing frequency of TE and TM mode. It is clear that by solving Eq. (4) one can have the perturbed band structure with band gap opening if the off-diagonal elements $\Delta_{TE,TM}$ and $\Delta_{TM,TE}$ are not zero. The eigenvalues and corresponding eigenmodes of Eq. (4) in the vicinity of $K$ and $K'$ point are

$$\omega'_\pm = \left(1 + \frac{\Delta_{TE,TE} + \Delta_{TM,TM}}{2}\right)\omega_D + \frac{\Delta_{TE,TE} - \Delta_{TM,TM}}{2} v_D \delta k \pm \sqrt{\left[\left(1 + \frac{\Delta_{TE,TE} + \Delta_{TM,TM}}{2}\right) v_D \delta k + \left(\frac{\Delta_{TE,TE} - \Delta_{TM,TM}}{2}\right)\omega_D\right]^2 + |\Delta_{TE,TM}|^2(\omega_D^2 - v_D^2 \delta k^2)}$$

, (5.1)

$$\psi^\uparrow = \begin{pmatrix} \Delta_{TE,TM} \\ -\frac{\Delta_{TE,TE}-\Delta_{TM,TM}}{2} + \sqrt{\left(\frac{\Delta_{TE,TE}-\Delta_{TM,TM}}{2}\right)^2 + |\Delta_{TE,TM}|^2} \end{pmatrix} \quad (5.2)$$

$$\psi^\downarrow = \begin{pmatrix} \frac{\Delta_{TE,TE}-\Delta_{TM,TM}}{2} - \sqrt{\left(\frac{\Delta_{TE,TE}-\Delta_{TM,TM}}{2}\right)^2 + |\Delta_{TE,TM}|^2} \\ \Delta_{TM,TE} \end{pmatrix}, \quad (5.3)$$

To evaluate $\Delta_{m,n}$, one can consider the parity of non-zero components of unperturbed TE and TM eigenfields. $H_{x,\text{TE}}(x, y, -z) = -H_{x,\text{TE}}(x, y, z)$ and $H_{y,\text{TE}}(x, y, -z) = -H_{y,\text{TE}}(x, y, z)$. $H_{x,TM}(x, y, -z) = +H_{x,TM}(x, y, z)$ and $H_{y,TM}(x, y, -z) = +H_{y,TM}(x, y, z)$.

$$\Delta_{TE,TE} = -\int_{\Delta V}(\epsilon_0|\boldsymbol{E}_{TE}|^2 - \mu_0|\boldsymbol{H}_{TE}|^2)dV, \quad (6.1)$$

$$\Delta_{TM,TM} = -\int_{\Delta V}(\epsilon_0|\boldsymbol{E}_{TM}|^2 - \mu_0|\boldsymbol{H}_{TM}|^2)dV, \quad (6.2)$$

$$\Delta_{TE,TM} = \Delta^*_{TM,TE} = \int_{\Delta V}\mu_0(\boldsymbol{H}^*_{TE} \cdot \boldsymbol{H}_{TM})dV, \quad (6.3)$$

Note that by introducing the asymmetric deformation $\Delta V$ with respect to the mirror symmetry $\sigma_z$, one can have finite $\Delta_{TE,TM}$ with non-vanishing bi-anisotropy. According to Eq. (5.1), the quadratic term in the square root thus exists, and the band gap opens. On the contrary, if $\Delta_{TE,TM} = 0$, the square root can be reduced to the linear function of $\delta k$, and the band gap fail to open. Because of the interaction between TE and TM mode, the degeneracy at $K$ and $K'$ point is lifted, and the eigenmodes [Eq. (5.2) and (5.3)] with distinct hybridization of the unperturbed TE and TM mode are introduced as spin-up ($\psi^\uparrow$) and spin-down ($\psi^\downarrow$) mode in analogy to their condensed matter counterparts. The effect of the bianisotropy can be effectively described by introducing a mass term of the form $\zeta \hat{s}_z \hat{\tau}_z \hat{\sigma}_z$ [14] into the effective Hamiltonian [Eq. (2)] which removes the internal U(1) polarization/spin symmetry resulting in new hybridized TE and TM modes.

The key manifestation of the topologically nontrivial nature of the spin-degenerate bi-anisotropic systems is the presence of the edge modes at the interface (domain wall) between two crystals with opposite bi-anisotropy [14]. From the symmetry considerations [25] it follows that such reversal of the bi-anisotropy can be achieved by applying the $\sigma_z$ (mirror image) operation to the original BMW, i.e. by opening the gap at the opposite end of the metallic cylinders. Therefore, the bi-anisotropic domain wall can be created by interfacing two BMWs with gaps orientation switching from upward to downward, as schematically shown in Fig. 2(a). It will be shown that such domain wall supports the "spin-locked" edge states with the phase relation between TE and TM mode locked to the direction of propagation along the interface. The latter property is the key feature which endows these modes with topological protection, i.e. robustness against different classes of structural imperfections.

First, to confirm that the domain wall between two BMWs supports the edge states within the complete band gap induced by the bi-anisotropy, we performed first principle PBS calculations (COMSOL Multiphysics) for the supercell consists of 60 unit cells with the domain wall at the supercell center [the flip in the gap position as is schematically shown in Fig. 2(a)]. The PBS shown in Fig. 2(b) spans the entire region of 1D Brillouin zone of the supercell. It shows that the supercell supports multiple modes propagating in the bulk of the BMWs (shown by black dotted lines). In addition, the PBS also reveals presence of the two modes within the band gap region (shown by blue lines). Inspection of the field profiles of these modes [Fig. 2(b)] shows that they represent edge modes localized to the domain wall. The dispersion lines of the edge modes run through the entire band gap region thus connecting the upper and lower bands of the bulk modes and have linear dependence near the $K$ and $K'$ points. Fig. 2(c) shows the electromagnetic energy density calculated in the openings between the cylinders and the metal plates. Because the edge modes are specific hybridization of the original TE and TM mode and the phase relation between the modes are locked to their propagation direction, the polarizations of forward and backward propagating mode can be distinguished. In Fig. 2(c), one can see a unique rotating pattern of the Poynting vectors winding around the cylinders' axes inside the gaps. The winding direction of the Poynting vector reverses across the domain wall thus giving rise to the directional net electromagnetic energy flow parallel to the domain wall. For instance, the mode exhibiting counter-clockwise rotation in the lower BMW and clockwise rotation in the upper BMW, which will be referred to as "spin-up" edge mode ($\psi^\uparrow$), propagates in the positive (forward) direction, as indicated by the top-right inset in Fig. 2(c). For the "spin-down" mode ($\psi^\downarrow$), shown in the bottom-right inset in Fig. 2(c), the direction of rotation in both top and bottom BMWs reverses along with the propagation direction of the mode.

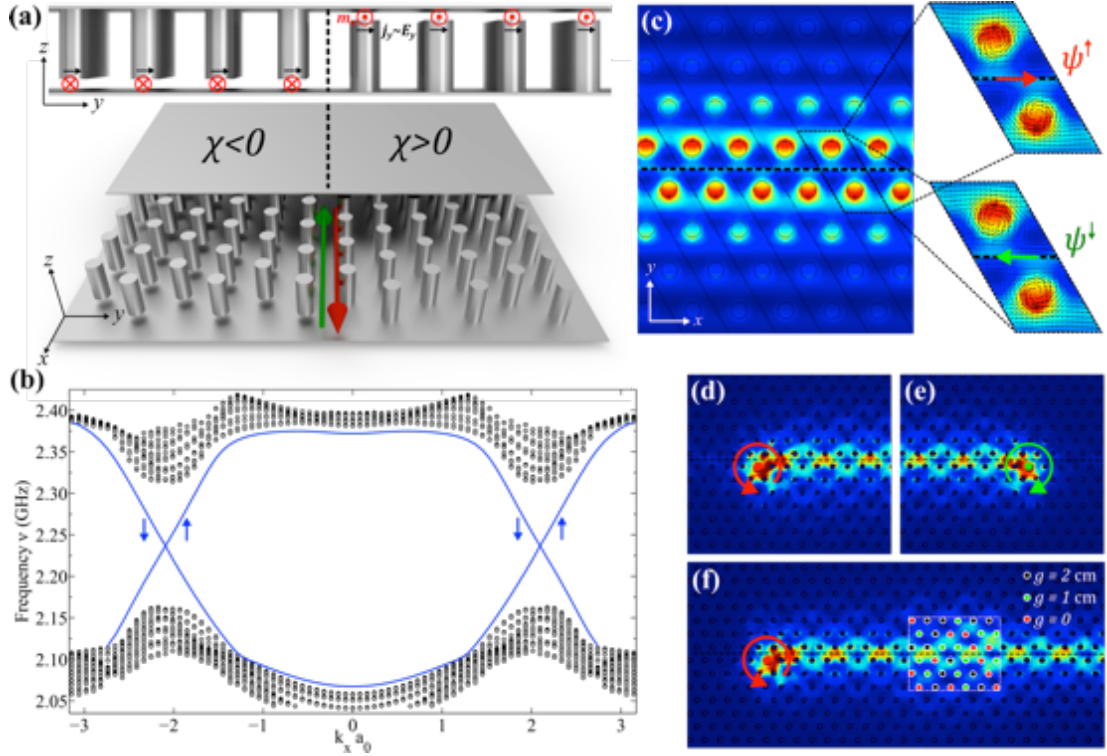

**Figure 2 (Color online).** (a) The top inset shows the side view of the domain wall between two BMWs with reversed bianisotropy at the center. The bottom inset is the 3D view of the domain wall with part of top metal plate removed to see the structure below. $\chi$ indicates the sign of bi-anisotropy in each domain (b) is the corresponding PBS of the 1D supercell formed by two BMWs. In (b) black dots correspond to bulk modes while blue lines correspond to edge modes confined to the domain wall and arrows show effective spin of the edge modes. (c) Electromagnetic energy density (color) and Poynting flux (black arrows in the inset with enlarged region) of the edge states for the two Kramer's partner modes. Spin-up and spin-down modes at K and K' points are indicated by red and green arrows, respectively. The fields are shown in the middle of the gaps between the cylinder and the metal plate corresponding to two different cut planes in upper and lower crystals. (d, e) Directional excitation of the spin-up and spin-down edge mode by the matching (rotating) local field pattern (clockwise and counter-clockwise rotating magnetic dipoles placed in the gap of the lower crystal, respectively). The structure parameters are the same as in Fig. 1 and $g = 1.5$ cm. (f) Reflectionless propagation of the spin-up edge mode through the 2D disordered region with randomized gap sizes. Gaps of different size are shown by circles of different color, as indicated in the legend and outside the disordered region $g = 2$ cm.

To further demonstrate that the edge modes confined to the domain wall are indeed spin-locked, we performed driven COMSOL Multiphysics simulations of the BMW consists of 20x32 unit cells (with the domain wall between $10^{th}$ and $11^{th}$ rows in $y$-direcion) and excited the modes by the matching rotating local filed profiles. Two cross-polarized (one along $x$ direction and one along $y$ direction) magnetic dipoles were placed at the center of the opening between one cylinder and the metal plate right next to the domain wall on its lower side [as indicated by red and green dots in Fig. 2(c)]. The $x$ and $y$ magnetic dipoles were driven with the relative phase of +90 and -90 degrees such that the clockwise or counter-clockwise magnetic dipolar patterns were produced. As can be seen from Fig. 2(d, e), such counter-rotating magnetic dipoles give rise to excitation of two counter-propagating edge modes thus evidencing spin-locked character of the edge modes. The distinctive polarizations of the modes enable us to directionally excite the edge modes by local (point-wise) source, which is not otherwise possible [26].

The edge modes of the bi-anisotropic domain wall of spin-degenerate ($\epsilon$-$\mu$-matched) metacrystals have been shown to be robust to various classes of structural imperfection such as sharp edges and disorders [14]. Therefore one can expect similar robustness of the edge modes supported by the domain wall of the BMWs considered here. However, because in the BMW case the engineering of the spin degree of freedom is largely reliant on the band structure, we cannot expect robustness against perturbations resulting in the removal of the degeneracy between the TE and TM modes near the Dirac point. One of such perturbation not preserving the "synthetic" spin degree of freedom of BMW is the randomized displacement of the cylinders in the $xy$-plane. It was confirmed by numerical calculations (not shown) that the edge modes do backscatter from the region with displaced. However, we also found by direct driven COMSOL simulations that the edge modes will remain topologically protected against a wide class of perturbations with preserving the spin of the BMW. As long as the spin property exists in the entire propagation path, the edge mode remains robust regardless of spatial inhomogeneity. Fig. 2(f) shows that a driven spin-up mode passes through a disorder region seamlessly. The disorder region (consisting of three different gap sizes randomly displaced) introduces an inhomogeneous distribution of the bi-anisotropy, but noticing that the upper domain always has gaps near the top metal plate and the lower domain has gaps near the bottom plate. The inhomogeneous bi-anisotropy cannot lift the edge modes because the local PBS of the upper domain is topologically non-trivial and has different topology from the local PBS of the lower domain. The local PBS of the upper domain thus cannot be continuously deformed to of the lower domain without closing the band gap. The Dirac point for the PBS of the edge modes is guaranteed as long as the bi-anisotropy flips sign across the domain wall.

The robustness of the edge mode against the other two types of disorder is shown in Fig. 3 and 4. We make a comparison between the non-trivial topological edge mode and a trivial-gapped [TM ($E_z$, $H_x$, $H_y$)] photonic defect mode. The schematic of the domain wall geometry for the defect mode is shown in Fig. 3(a). It consists of semicircular metallic rods with flip of the opening direction across the domain wall. By breaking the in-plane spatial inversion symmetry, the original PBS with TM Dirac point (PBS of circular-metallic-cylinder geometry) is lifted. A complete band gap is formed. The 1D PBS of the domain wall structure is also calculated [shown in Fig. 3(b)]. In this particular geometry there is only one photonic defect mode exists in the band gap region, which makes it a perfect candidate for the comparison with topological edge mode. We first compare the "scattering" of the edge mode and the defect mode by a "cavity" placed on the path of the modes. The cavity represents an extended spatial region with local PBS having Dirac cones. For the defect mode, the cavity is made of metallic cylinders with diameter

$4.45cm$ arranged as hexagonal lattice as indicated with red circles in Fig. 3(c). Similarly for the edge mode, the cavity consists of the metallic cylinders with diameter $3.45cm$ and gaps ($g = 0.5cm$) both on top and bottom of the cylinders. Since the gaps are introduced symmetrically, there is no bi-anisotropy. The positions of the cylinders with symmetric gaps are indicated in Fig. 3(e) by red circles. Because these cavity regions have no broken in-plane spatial inversion symmetry or bi-anisotropy, it supports degenerate modes (PBS with Dirac points) propagating in the bulk. Fig. 3(d) shows the transmission spectrum for the defect mode with frequency within the band gap region. Several sharp peaks are observed as a result of backscattering inside the cavity region. Because the forward propagating defect modes can freely scatter back once it hits disorders, the electromagnetic energy can build up in the cavity. Occasionally if the driving frequency hits a resonance frequency of the cavity, one can get resonant tunneling as shown in the right inset of Fig. 3(c). By comparing the transmission spectra [Fig. 3(d, f)] one can see that the photonic defect mode can only achieve 90% transmission for 9.4% of the frequency range in the band gap, while the topological edge mode achieve 90% transmission for 100% of the frequency range in the band gap.

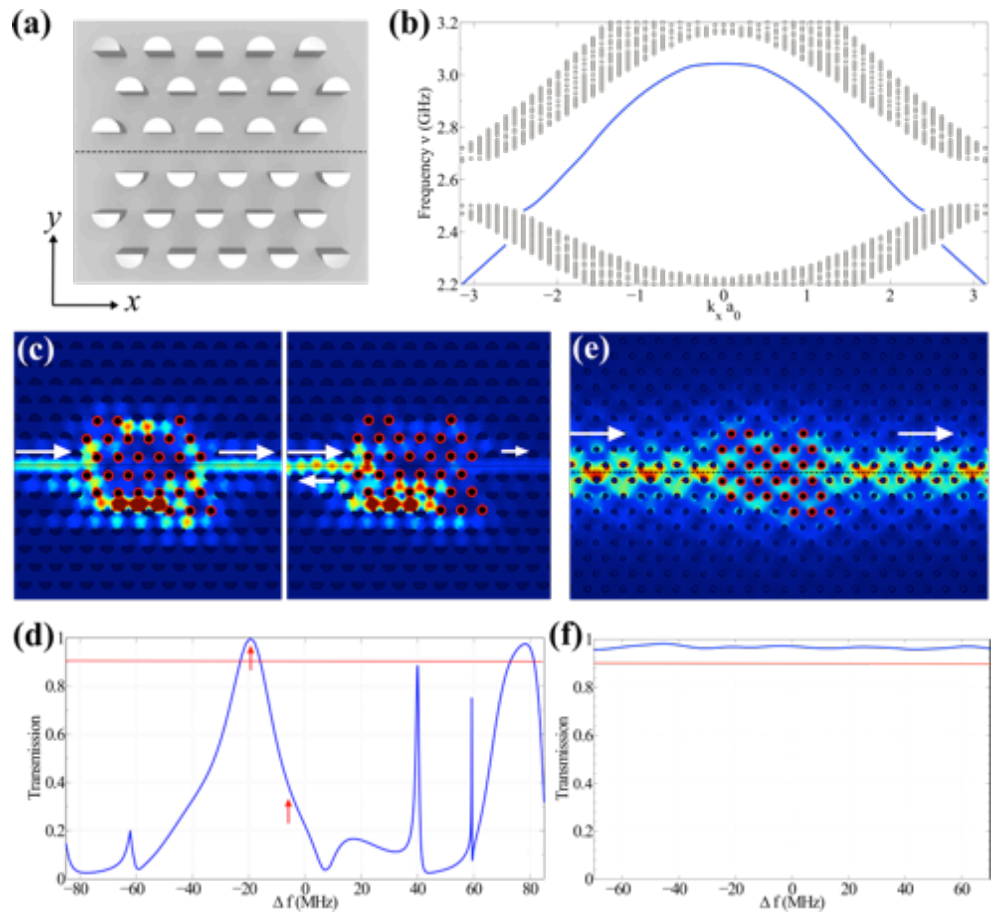

**Figure 3 (Color online).** (a) Schematic of the domain wall between two trivial-gapped photonic crystals with reversed opening directions of metallic semicircular rods. (b) is the corresponding PBS of the 1D supercell formed by trivial-gapped photonic crystals. In (b) black dots correspond to bulk modes while blue lines correspond to the photonic defect mode confined to the domain wall. (c) The left inset shows the energy density of resonant-tunneling photonic defect mode through the cavity formed by circular metallic cylinders. Red circles indicate locations of the cylinders. The frequency of the defect mode indicated by red arrow at the peak in (d). The right inset shows the case of off-resonance. (d) Transmission spectrum of the defect mode with the frequency range in the band gap (centered at 2.59 GHz). Red line at $T = 0.9$ guides the eye. (e) Reflectionless propagation of the edge mode through the cavity formed by gapless ($g = 0$, non-bi-anisotropic MW). (f) Transmission spectrum of the edge mode with the frequency range in the band gap (centered at 2.24 GHz).

The other class of even more strong perturbations is demonstrated in Fig. 4. A sharp bends of the domain wall forms a zigzag-shape channel. The defect mode has the similar transport behavior as it is in the cavity study. Fig. 4(b) shows the 100% transmission for the defect mode can be achieved if the driving frequency is tuned to be a resonance frequency of the middle part of the zigzag channel. In Fig. 4(c), the topological edge mode is excited at the top horizontal domain wall seamlessly changes its propagation direction by 120 degree and continues to follow the domain wall without back-reflection. The same behavior is observed at the second bending of the domain wall where the mode restores its original propagation direction without back reflection. We again compare the transmission spectrum of the defect mode and the edge mode. The defect mode achieve 90% transmission for only 5.6% of the frequency range in the band gap, while the topological edge mode achieve 90% transmission for 54% of the frequency range in the band gap.

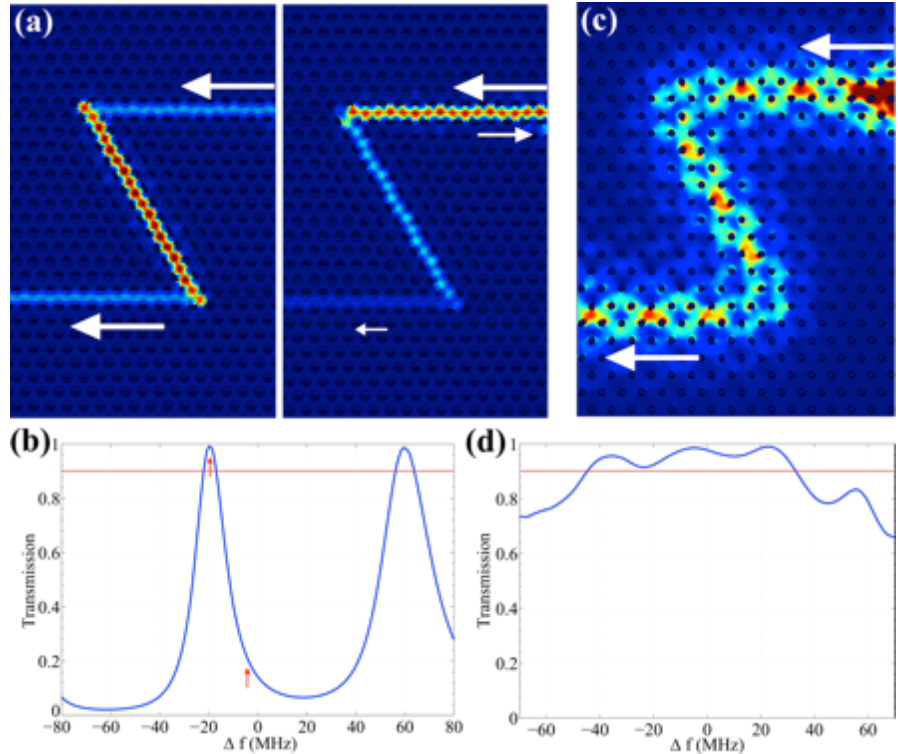

**Figure 4 (Color online).** (a) The left inset shows the energy density of resonant-tunneling photonic defect mode through a zigzag shaped channel between two trivial-gapped photonic crystal. The frequency of the defect mode indicated by red arrow at the peak in (b). The right inset shows the case of off-resonance. (b) Transmission spectrum

of the defect mode with the frequency range in the band gap (centered at 2.59 GHz). Red line at $T = 0.9$ guides the eye. (c) Reflectionless routing of the edge mode propagating along the zigzag shaped interface between two BMWs. (d) Transmission spectrum of the edge mode with the frequency range in the band gap (centered at 2.24 GHz).

It is worth to stress that the BMW structure we are proposing in this letter doesn't rely on special homogenized material responses. Spin degeneracy is achieved (degenerate Dirac cones of PBS for TE and TM mode) by the symmetries of this particular geometry itself. Since we don't need the identical and strong anisotropic $\epsilon$ and $\mu$ tensor to emulate the spin degree of freedoms [14] and the strong bi-anisotropic response can also be introduced in a non-resonant way, the edge mode frequency can be far away from any of the resonance of electric or magnetic dipole moments. This gives a tremendous advantage for BMW that it is greatly scalable from radiofrequencies to infrared. As one starts to approach infrared regime, the resistive loss in metal cannot be ignored. To examine the effect of ohmic loss in metal on the edge modes we performed calculation with the conductivity of aluminum ($\sigma = 3.5 \times 10^7$ S/m) and found that the propagation length $l = 0.5 \times Im(k^{-1})$ of the edge modes is as large $3 \times 10^5$ periods ($l = 3 \times 10^6$), i.e. is literally infinite. Moreover, by scaling down the dimensions of the structure by a factor of $2 \times 10^4$ and the adjusting geometric parameters of the structure to take into account for the plasmonic response, we designed BMW made of gold and operation at $5 \mu$m. In this case with loss in the gold considered we found the propagation length as long as 50 periods.

To summarize, we introduced a novel design of bi-anisotropic meta-waveguide exhibiting topologically nontrivial photonic modes. Domain walls between such waveguides with reversed bi-anisotropy were shown to support guided edge modes with their spin locked to the propagation direction which ensured robustness of the modes against various classes of structural imperfections. Such robustness was confirmed by the full-vectorial numerical simulation for the cases of cavities, sharp bends and disordered regions placed on the edge modes propagation path. Broadband perfect transport behaviors of the topological edge modes are also demonstrated. We believe that this unique property of the edge modes makes them perfect candidates for various applications from radiofrequencies to infrared where the directional excitation and reflectionless routing of the electromagnetic waves is required.